# Transformer-Hypernetwork-Controlled Deep-Unfolded Phase-Aware Channel Estimation Refinement for Phase-Drift-Robust Backscatter Links

Hanyeol Ryu, Nohgyeom Ha, and Sangkil Kim, *Senior Member, IEEE*

***Abstract*—This paper proposes a transformer-hypernetwork-controlled deep-unfolded phase-aware channel estimation refinement (THUNDER) for phase-drifting backscatter links. Residual carrier-phase drift across the pilot block renders the backscattered observation phase-nonstationary, and a closed-form phase-aware channel estimation (PACE) compensates only the first-order phase component, leaving a deterministic high signal-to-noise ratio (SNR) error floor. THUNDER suppresses this floor by initializing from PACE and refining the estimate through unfolded Gauss–Newton steps on the exact phase-exponential model. A transformer extracts pilot-wide phase context, and a hypernetwork generates bounded controls and pilot-reliability weights. Evaluations show an 8.9 dB normalized mean square error gain over the strongest learning-based channel estimation baseline.**

## I. INTRODUCTION

BACKSCATTER communication enables battery-free Internet-of-Things (IoT) connectivity by allowing a passive tag to reflect an incident carrier rather than generate its own [1, 2]. In a monostatic link, residual carrier-phase drift accumulates across the pilot block and varies between frames, making the cascaded-channel observation phase-nonstationary and invalidating the constant-phase assumption of conventional channel estimation (CE) [3].

A closed-form phase-aware channel estimation (PACE) jointly initializes the effective channel and residual phase slope through a first-order lifted model [3]. However, by truncating the phase exponential after its linear term, this model leaves a deterministic residual that grows with the pilot phase span and induces a high signal-to-noise ratio (SNR) error floor.

Deep learning (DL) methods assist CE by learning nonlinear channel features from pilot observations [4]–[7]. Existing DL-based CE can be broadly categorized into direct pilot-to-channel mapping and model-assisted estimation, where learned modules refine or control an analytical estimator. In phase-drifting backscatter links, direct mapping leaves the per-frame phase-exponential observation unconstrained, whereas model-assisted updates with fixed or globally shared controls remain blind to the conditioning and reliability of each pilot block. A remaining gap is a PACE-preserving estimator that retains the closed-form channel-and-phase state, refines it on the exact phase-exponential observation, and adapts bounded update controls and pilot weights from the complete pilot sequence.

To address this gap, this paper proposes the transformer-hypernetwork-controlled deep-unfolded PACE refinement (THUNDER). The PACE state initializes a finite-depth Gauss–Newton (GN) recursion on the exact phase-exponential observation model. The transformer encodes the pilot context, from which the hypernetwork and a token-wise head generate step-dependent GN controls and pilot-reliability weights, respectively. Bounded control mappings, trust-region projection, and a residual guard constrain the refinement without replacing the model-based channel-and-phase state, thereby extending PACE beyond the first-order operating regime. The key contributions are as follows:

- An exact-model deep-unfolded GN refinement is developed by initializing from PACE and iterating on the phase-exponential observation, suppressing the deterministic error floor induced by the first-order approximation.
- A transformer-hypernetwork controller is introduced to convert pilot-wide context into bounded step size, GN damping, loading, and pilot-reliability weights, enabling frame-dependent refinement without replacing the channel-and-phase state.
- A feature-extraction transfer protocol is evaluated under source-to-target mismatch by freezing the transformer and adapting only the hypernetwork and reliability head.

## II. SYSTEM MODEL

### A. Phase-Nonstationary Backscattered Pilot Observation

Consider the monostatic single-input single-output (SISO) backscatter link shown in Fig. 1, in which a co-located reader illuminates a load-switching tag and estimates the cascaded channel from the known reflected pilot block. The effective channel is defined as $h_{eq} \triangleq h_{tt}h_{tr}$, where $h_{tt}$ and $h_{tr}$ denote the transmitter-to-tag and tag-to-receiver channel coefficients, respectively. Each hop $h_i$, $i \in \{tt, tr\}$, follows a Rician distribution with $K$-factor $K_i$ and average power $\Omega_i = \mathbb{E}[|h_i|^2]$, and their product $h_{eq}$ follows the cascaded double-Rician fading channel model in [8].

This work was supported by the Nano & Material Technology Development Program through the National Research Foundation of Korea (NRF) funded by Ministry of Science and ICT (RS-2025-02304267). (*Corresponding author: Sangkil Kim*).

Hanyeol Ryu and Sangkil Kim are with the Department of Electronics Engineering, Pusan National University, Busan 46241, Republic of Korea (e-mail: ksangkil3@pusan.ac.kr).

Nohgyeom Ha is with the Department of Satellite System, Hanwha Systems, Seongnam-si, 13591, Republic of Korea.

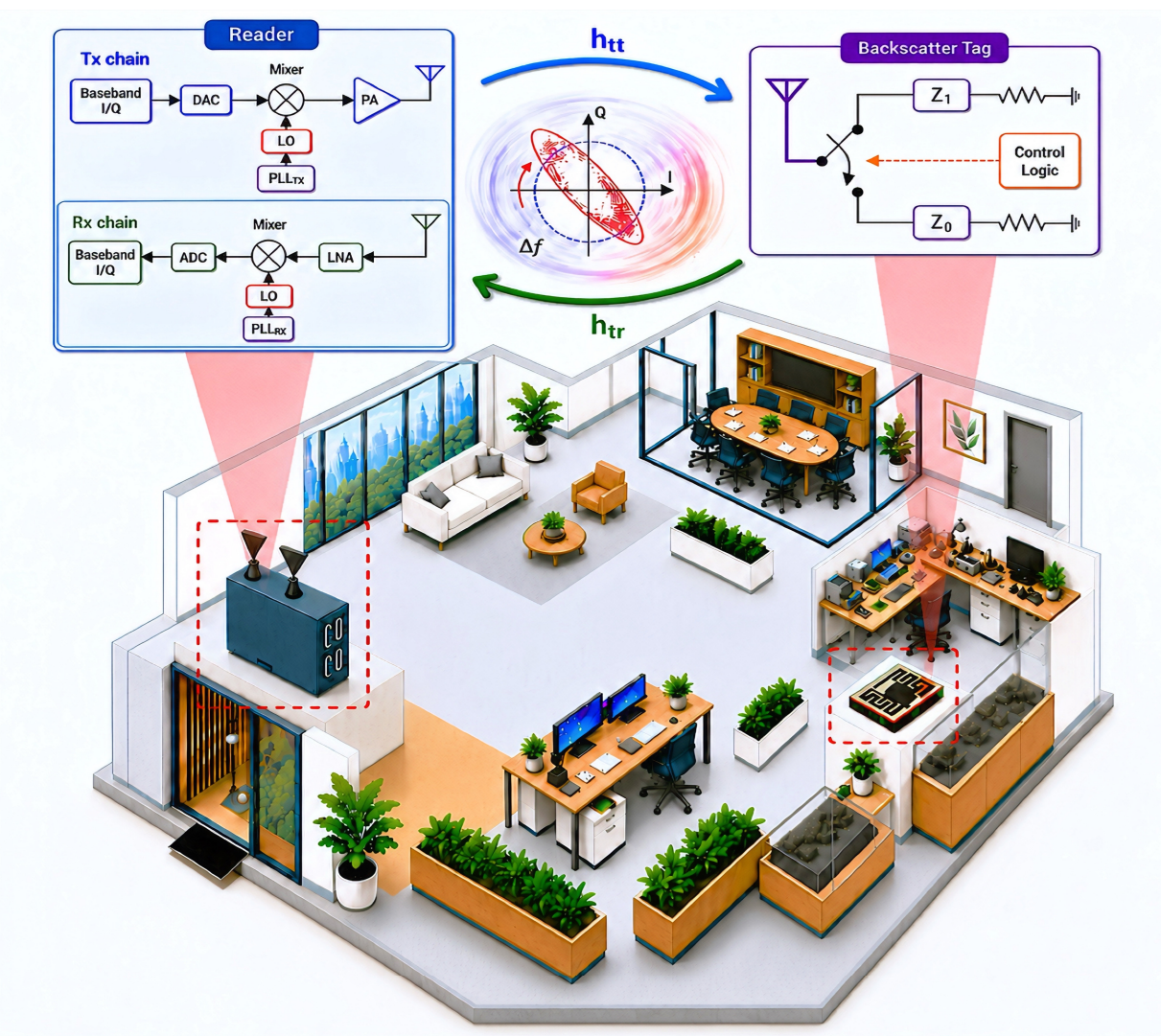

Fig. 1. Phase-nonstationary monostatic indoor backscatter environment, where residual carrier-phase drift distorts the pilot observation.

Let $\mathcal{P}$ denote the pilot index set, with $x_n$ and $y_n$ representing the transmitted and received pilot samples at index $n$, respectively ($|\mathcal{P}| = 30$). The channel is modeled as flat fading over each pilot block; hence $h_{eq}$ remains constant for $n \in \mathcal{P}$ but may vary between frames. The residual carrier-phase drift remaining after coarse carrier compensation is modeled by the residual phase slope $\phi_1$, which induces a linear phase progression across the pilot indices. Retaining this progression in exponential form, the exact pilot observation is defined as

$$y_n = h_{eq} x_n e^{j\phi_1 n} + w_n, \qquad n \in \mathcal{P}, \tag{1}$$

where $\phi_1$ is the residual phase slope in rad/sample, and $w_n$ denotes the effective post-processed noise sample. The pilot phase span is defined as $\Phi_s = |\phi_1|(n_{\max} - n_{\min})$, where $n_{\max}$ and $n_{\min}$ are the largest and smallest indices in $\mathcal{P}$, respectively. As $\Phi_s$ increases, the accumulated phase rotation term in (1) increasingly violates the pilot-coherence assumption underlying phase-stationary estimators.

### B. Phase-Aware Channel Estimation Initialization under First-Order Model Mismatch

The PACE approximates the residual phase exponential as $e^{j\phi_1 n} \approx 1 + j\phi_1 n$, yielding the lifted pilot observation

$$y_n \approx \theta_0 x_n + \theta_1 n x_n + w_n. \tag{2}$$

Here, $\theta_0 = h_{eq}$ denotes the zeroth-order channel coefficient, while $\theta_1 = j\phi_1 h_{eq}$ captures its first-order coupling with the residual phase slope [3]. The closed-form least-squares solution of (2) provides $\hat{\theta}_0$ and $\hat{\theta}_1$, with $\hat{h}_{eq}^{(0)} = \hat{\theta}_0$. The residual phase-slope initializer follows from regularized normalization of $\hat{\theta}_1$ by $\hat{\theta}_0$, $\hat{\phi}_1^{(0)} = \Im\{\hat{\theta}_1 \hat{\theta}_0^* / (|\hat{\theta}_0|^2 + \epsilon)\}$, where $\epsilon$ prevents unstable normalization when $|\hat{\theta}_0|$ is small. The resulting channel and phase-slope estimates form the initial state

$$\boldsymbol{\eta}^{(0)} = \left[\Re\left\{\hat{h}_{eq}^{(0)}\right\}, \Im\left\{\hat{h}_{eq}^{(0)}\right\}, \hat{\phi}_1^{(0)}\right]^T. \tag{3}$$

Although $\boldsymbol{\eta}^{(0)}$ provides a closed-form PACE anchor for the channel and residual phase slope, the lifted model leaves the deterministic residual $m_n = h_{eq} x_n (e^{j\phi_1 n} - 1 - j\phi_1 n)$. For real $\phi_1 n$, $|m_n| \leq (|h_{eq} x_n||\phi_1 n|^2)/2$; hence, this mismatch grows quadratically with the phase coordinate and is independent of the additive-noise variance. Increasing the SNR reduces the stochastic estimation error but leaves $m_n$ unchanged, causing the PACE error to saturate once $\Phi_s$ exceeds the first-order operating regime.

## III. Proposed THUNDER Channel Estimator

The deterministic residual, $m_n$, lies outside the lifted model in (2), and therefore remains beyond elimination within the first-order representation. THUNDER therefore retains $\boldsymbol{\eta}^{(0)}$ as a PACE anchor and unfolds exact-model updates conditioned on pilot context. The overall architecture of THUNDER is shown in Fig. 2.

### A. Exact-Model Deep-Unfolded Gauss-Newton Refinement

At refinement step $k$, the current channel and phase-slope estimates are collected as

$$\boldsymbol{\eta}^{(k)} = \left[\Re\left\{\hat{h}_{eq}^{(k)}\right\}, \Im\left\{\hat{h}_{eq}^{(k)}\right\}, \hat{\phi}_1^{(k)}\right]^T. \tag{4}$$

Starting from (3), the exact-model prediction is $\hat{y}_n^{(k)} = \hat{h}_{eq}^{(k)} x_n e^{j\hat{\phi}_1^{(k)} n}$, and the residual is $r_n^{(k)} = y_n - \hat{y}_n^{(k)}$. Stacking $r_n^{(k)}$ over $n \in \mathcal{P}$ gives $\mathbf{r}_k \in \mathbb{C}^{|\mathcal{P}|}$. The Jacobian $\mathbf{J}_k \in \mathbb{C}^{|\mathcal{P}| \times 3}$ is evaluated with respect to the real-valued state in (4). In particular, the row of $\mathbf{J}_k$ associated with pilot index $n$ is

$$\mathbf{j}_n^{(k)} = \left[x_n e^{j\hat{\phi}_1^{(k)} n}, j x_n e^{j\hat{\phi}_1^{(k)} n}, j n \hat{h}_{eq}^{(k)} x_n e^{j\hat{\phi}_1^{(k)} n}\right]. \tag{5}$$

The pilot-reliability head assigns a bounded weight to each pilot sample at step $k$. The weights are normalized over $\mathcal{P}$ and collected in the diagonal matrix $\mathbf{W}_k$. This normalization changes the relative pilot contributions without changing the average scale of the normal equation. The weighted normal matrix and vector are then $\mathbf{H}_k = \Re\{\mathbf{J}_k^H \mathbf{W}_k \mathbf{J}_k\}$, and $\mathbf{g}_k = \Re\{\mathbf{J}_k^H \mathbf{W}_k \mathbf{r}_k\}$, yielding the damped-GN increment

$$\Delta\boldsymbol{\eta}^{(k)} = \left(\mathbf{H}_k + \lambda_k \mathbf{I} + \mathrm{diag}(\mathbf{p}_k)\right)^{-1} \mathbf{g}_k. \tag{6}$$

Here, $\lambda_k > 0$ adds isotropic damping, while $\mathbf{p}_k \in \mathbb{R}_+^3$ provides coordinate-wise loading for the two channel components and phase slope. These controls regularize the global conditioning and the state-direction scale imbalance separately. The resulting increment is scaled and trust-region projected as

$$\boldsymbol{\eta}^{(k+1)} = \boldsymbol{\eta}^{(k)} + \alpha_k \Pi_{\mathcal{T}}\left(\Delta\boldsymbol{\eta}^{(k)}\right), \tag{7}$$

where $\alpha_k$ is the step size, and $\Pi_{\mathcal{T}}(\cdot)$ clips the channel and phase-slope increment to $[-\tau_h, \tau_h]$ and $[-\tau_\phi, \tau_\phi]$, respectively. This projection prevents an ill-conditioned local normal equation from inducing an arbitrarily large one-step update. Let $\tilde{\boldsymbol{\eta}} = \boldsymbol{\eta}^{(L)}$ denote the candidate state after the $L$ unfolded updates. For a state $\boldsymbol{\eta} = [\eta_1, \eta_2, \eta_3]^T$, define the observable exact-model residual energy as $\mathcal{R}(\boldsymbol{\eta}) = |\mathcal{P}|^{-1} \sum_{n \in \mathcal{P}} |y_n - (\eta_1 + j\eta_2) x_n e^{j\eta_3 n}|^2$. The residual guard returns

$$\hat{\boldsymbol{\eta}} = \begin{cases} \tilde{\boldsymbol{\eta}}, & \mathcal{R}(\tilde{\boldsymbol{\eta}}) \leq (1 + \tau_g)\mathcal{R}(\boldsymbol{\eta}^{(0)}), \\ \boldsymbol{\eta}^{(0)}, & \text{otherwise}, \end{cases} \tag{8}$$

where $\tau_g \geq 0$ is the allowable residual-increase tolerance. Since $\mathcal{R}(\boldsymbol{\eta})$ depends only on the received pilots, known pilots, and pilot indices, the guard is available during inference without true-channel knowledge. The final estimates are $\hat{h}_{eq} = \hat{\eta}_1 + j\hat{\eta}_2$ and $\hat{\phi}_1 = \hat{\eta}_3$, from which the reconstructed pilot is $\hat{y}_n = \hat{h}_{eq} x_n e^{j\hat{\phi}_1 n}$.

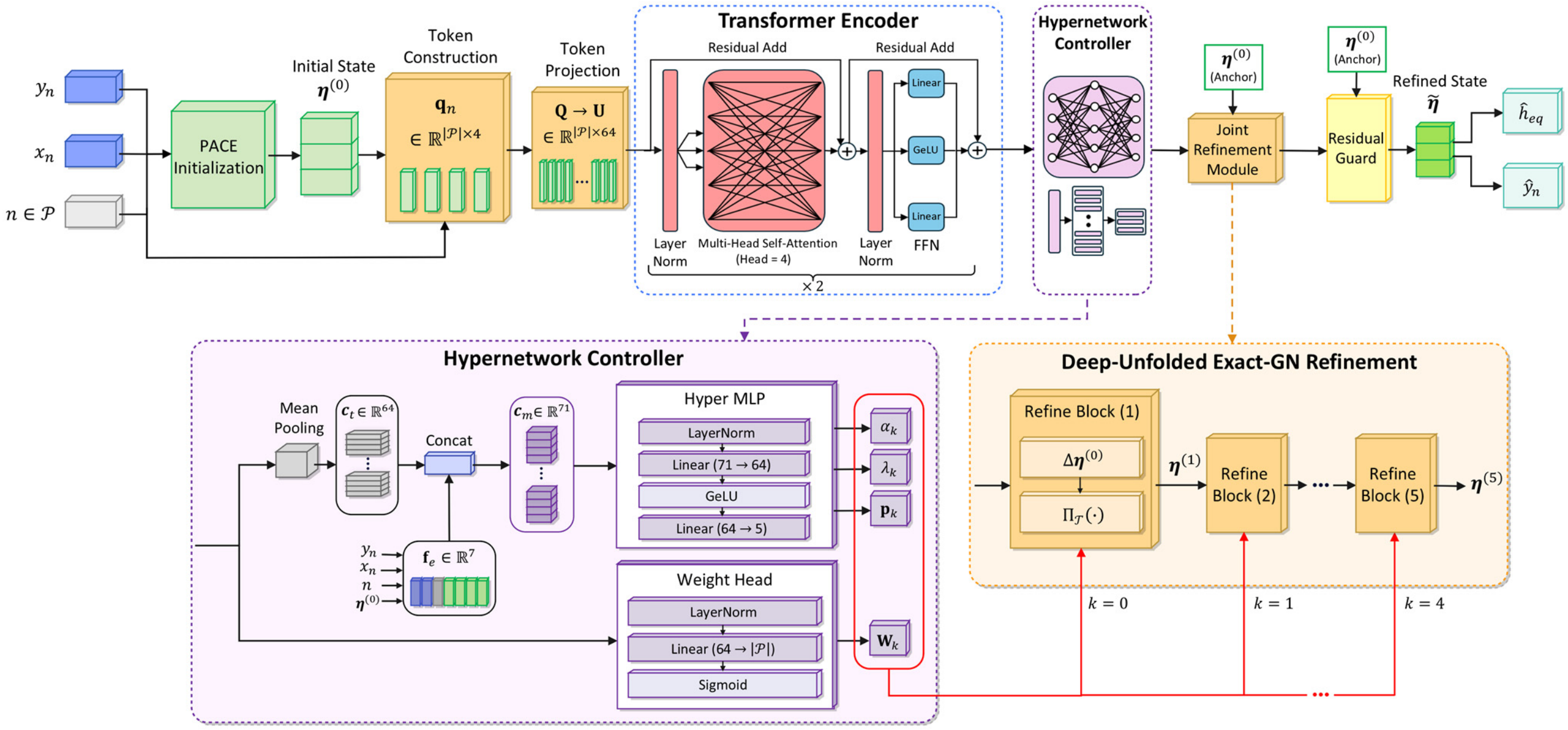

Fig. 2. The proposed THUNDER architecture for backscatter CE.

The exact-model residual and Jacobian keep the GN refinement tied to the physical observation model, whereas learning adjusts only $\alpha_k$, $\lambda_k$, $\mathbf{p}_k$, and $\mathbf{W}_k$. This separation removes the first-order restriction while preserving the physical interpretation of the model-based state.

### *B. Transformer-Hypernetwork Control*

Frame-dependent controls are required because phase drift and pilot reliability are block-level properties. A transformer encoder therefore forms pilot-wide context by relating each pilot token to all other tokens through self-attention [9]. Such pilot-wide contextualization is motivated by the residual phase slope being inferred from relative phase evolution across the pilot block, rather than from isolated pilot samples.

For pilot index $n$, define the pilot-matched observation as $\bar{z}_n = x_n^* y_n$, which is normalized by the pilot-energy scale $\rho$. The normalized pilot index $u_n$ encodes the relative position of sample $n$ within the pilot block, and $\zeta_n$ denotes a bounded version of the PACE phase coordinate $\hat{\phi}_1^{(0)} n/\pi$. The input token is

$$\mathbf{q}_n = \left[\frac{\Re\{\bar{z}_n\}}{\rho}, \frac{\Im\{\bar{z}_n\}}{\rho}, u_n, \zeta_n\right]^T. \quad (9)$$

Stacking all pilot tokens gives $\mathbf{Q} = [\mathbf{q}_n^T]_{n\in\mathcal{P}} \in \mathbb{R}^{|\mathcal{P}|\times 4}$. A learned token projection maps $\mathbf{Q}$ to $\mathbf{U} \in \mathbb{R}^{|\mathcal{P}|\times 64}$, and the transformer encoder converts $\mathbf{U}$ into the contextual token matrix $\mathbf{C} = [\mathbf{c}_n^T]_{n\in\mathcal{P}} \in \mathbb{R}^{|\mathcal{P}|\times 64}$. Mean pooling over the pilot dimension, followed by a context projection, yields the global pilot context $\mathbf{c}_t \in \mathbb{R}^{64}$.

The transformer output serves only as pilot context and does not directly regress the channel. A frame feature vector $\mathbf{f}_e \in \mathbb{R}^7$ describes the PACE channel direction and magnitude, initial phase slope, maximum phase coordinate, and residual-to-signal power ratios. The controller input is the concatenated vector $\mathbf{c}_m = [\mathbf{c}_t^T, \mathbf{f}_e^T]^T \in \mathbb{R}^{71}$.

For each unfolding step, the hypernetwork maps $\mathbf{c}_m$ to $\alpha_k$, $\lambda_k$, and $\mathbf{p}_k$. A separate token-wise linear head maps each contextual embedding $\mathbf{c}_n$ to one reliability logit per unfolding step; the logits associated with step $k$ form the pilot scores used to construct $\mathbf{W}_k$. Hence, the frame-level context controls the update scale and regularization, while the token-level context controls the relative contribution of each pilot observation. Bounded positive mappings constrain the generated controls around a damped-GN reference schedule, so the neural modules adapt the exact-model recursion rather than replace the model-based update [10].

### *C. Training and Transfer Adaptation*

The trainable modules are optimized end-to-end through the unfolded refinement. The loss combines mean effective-channel normalized mean square error (NMSE) $\mathcal{L}_h$, a tail-error term $\mathcal{L}_t$, a failure penalty $\mathcal{L}_f$, and a regularization term $\mathcal{L}_r$ for phase-slope error, pilot reconstruction, residual monotonicity, and control deviation. With fixed nonnegative coefficients $\beta_t$, $\beta_f$, and $\beta_r$, the training objective is

$$\mathcal{L} = \mathcal{L}_h + \beta_t\mathcal{L}_t + \beta_f\mathcal{L}_f + \beta_r\mathcal{L}_r. \quad (10)$$

The exact observation model, the closed-form PACE initialization rule, and the GN update equations remain fixed during training; only the parameters of the transformer encoder, hypernetwork, and reliability head are optimized.

After source-domain pretraining, four protocols are compared: source-only inference, target-only training, feature-extraction transfer, and full fine-tuning. Feature-extraction transfer freezes the transformer encoder and updates only the hypernetwork and reliability head, whereas full fine-tuning updates all trainable modules [11]. Across the target-domain protocols, the target-data budget and optimization schedule are matched. Validation-selected parameters are then fixed for inference on independently generated test realizations.

TABLE I
MAIN PARAMETERS OF THUNDER

| Parameter | Value |
|---|---|
| Model dimension | 64 |
| Attention heads / encoder layers | 4 / 2 |
| Feedforward dimension | 128 |
| Dropout | 0.03 |
| Optimizer | AdamW |
| Batch size | 128 |
| Source / target training steps | 2400 / 1400 |
| Source / target learning rate | $5 \times 10^{-4}$ / $1 \times 10^{-4}$ |
| Source / target $\Phi_s$ | 0°–80° / 0°–160° |
| Source / target $\rho_w$ | 0 / $\mathcal{U}(0, 0.8)$ |
| $K_{tt}$, $K_{tr}$ | i.i.d. $\mathcal{U}(0, 14)$ dB |
| $\Omega_{tt}$, $\Omega_{tr}$ | Unit normalized |
| Pilot length | 30 symbols |

## IV. NUMERICAL RESULTS

The evaluation is designed to verify whether THUNDER suppresses the deterministic error floor induced by the first-order approximation, rather than merely improving the average NMSE. All channels follow the cascaded double-Rician fading channel model, and the received pilots are generated from the exact pilot observation in (1). The NMSE of $h_{eq}$ is used as the accuracy metric, with the main settings listed in Table I.

### A. Simulation Settings

The per-hop Rician factors $K_{tt}$ and $K_{tr}$ are independently sampled from $\mathcal{U}(0,14)$ dB, while $\Omega_{tt} = \Omega_{tr} = 1$ fixes the average hop powers. This setting covers symmetric and asymmetric two-hop fading conditions without conflating the $K$-factor variation with an average-SNR shift. The target domain additionally includes temporally correlated additive noise following $w_n = \rho_w w_{n-1} + \sqrt{1-\rho_w^2}\, v_n$, where $\rho_w$ is the temporal correlation coefficient, $v_n \sim \mathcal{CN}(0, \sigma_w^2)$, and $\sigma_w^2$ denotes the noise variance specified by the SNR. The initial noise sample follows $w_0 \sim \mathcal{CN}(0, \sigma_w^2)$, so the recursion changes temporal correlation without changing the average noise power. The source domain uses the white-noise case $\rho_w = 0$, whereas $\rho_w$ is sampled independently for each target-domain realization.

Baselines include conventional phase-stationary CE, PACE in (3), fixed GN refinement on the exact model with pilot-independent controls, direct transformer regression, PACE-feature transformer regression, and an exact-model nonlinear least-squares (NLS) reference that fits (1) directly over $h_{eq}$ and $\phi_1$ without finite-order truncation or learning. This set separates the effects of phase-stationary CE, first-order phase modeling, exact-model recursion, pilot-dependent control, transformer-based regression with or without PACE features,

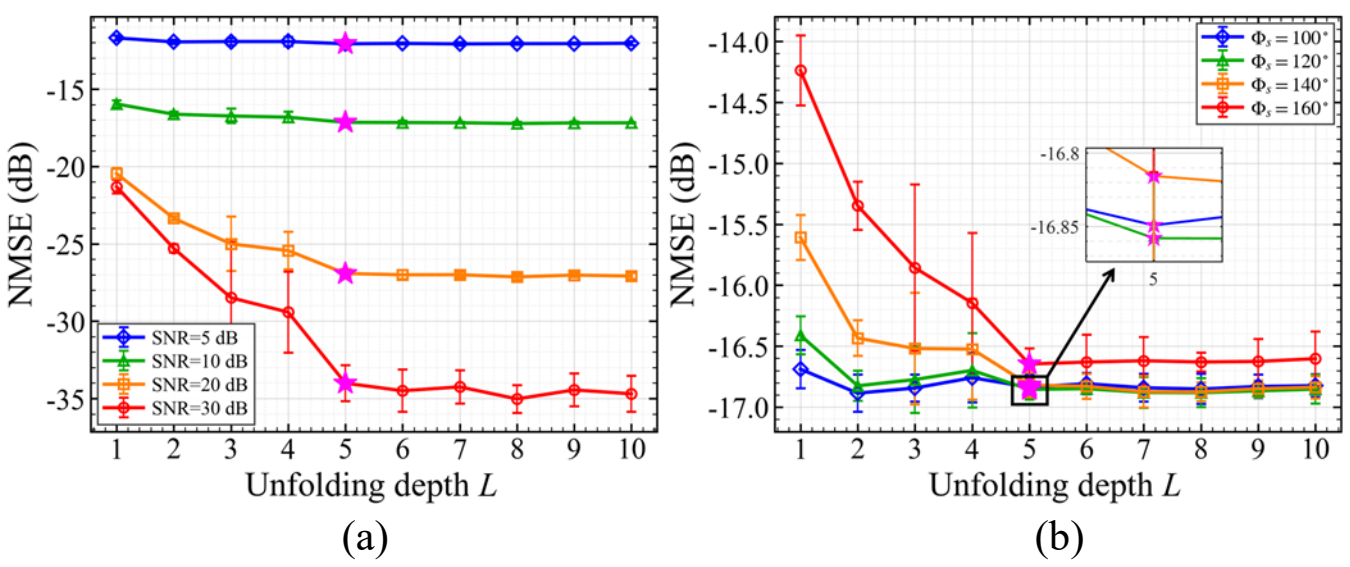

Fig. 3. Sensitivity of the unfolding depth $L$ over (a) SNR and (b) pilot phase span $\Phi_s$. Star markers ($L^{\star}$) indicate the depth selected under a 95% confidence level.

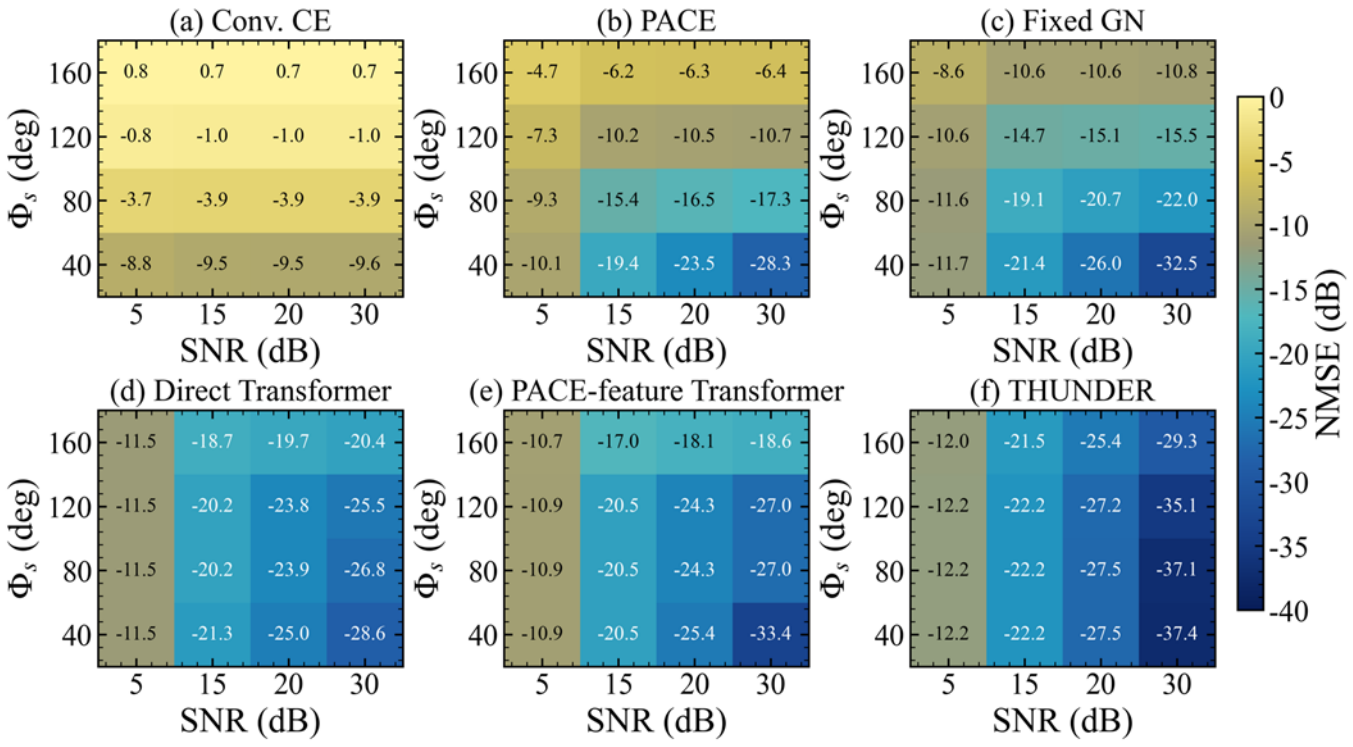

Fig. 4. NMSE over SNR and pilot phase span $\Phi_s$ for (a) conventional phase-stationary CE, (b) PACE, (c) fixed GN, (d) direct transformer, (e) PACE-feature transformer, and (f) THUNDER.

and the asymptotic accuracy attainable without any truncation or learned control.

The unfolding depth $L$ was selected by validation to balance accuracy and complexity. The selected depth was $L^{\star} = 5$, the smallest depth statistically indistinguishable from the validation-best depth at the 95% confidence level. As shown in Fig. 3(a) and (b), the NMSE saturates near $L = 5$, and deeper unfolding provides less than 0.5 dB additional gain. Since each step beyond $L^{\star}$ increases the GN-refinement cost by about $1/L^{\star} \approx 20\%$, this marginal gain offers little benefit relative to the added cost.

### B. Performance Evaluation

The heatmaps in Fig. 4 show that THUNDER yields the largest gain in the high-SNR and large-$\Phi_s$ region, where deterministic model mismatch dominates. At SNR = 30 dB and $\Phi_s$ = 160°, THUNDER improves the NMSE by 22.9 dB over PACE, 18.5 dB over fixed GN, and 8.9 dB over direct transformer. This confirms that the gain comes from exact-model refinement combined with pilot-context-dependent control.

To separate Taylor-order effects from exact-model refinement, an approximation-order comparison is performed at SNR = 30 dB over $\Phi_s$. As shown in Fig. 5, higher-order PACE reduces the mismatch at moderate spans but still degrades at large $\Phi_s$, because the phase exponential remains finitely truncated. THUNDER remains much closer to the exact-model NLS reference than finite-order PACE, improving over the third-order PACE by about 15 dB at $\Phi_s$ = 160°. This result indicates that the main gain comes from exact phase-exponential refinement rather than from simply increasing the Taylor order.

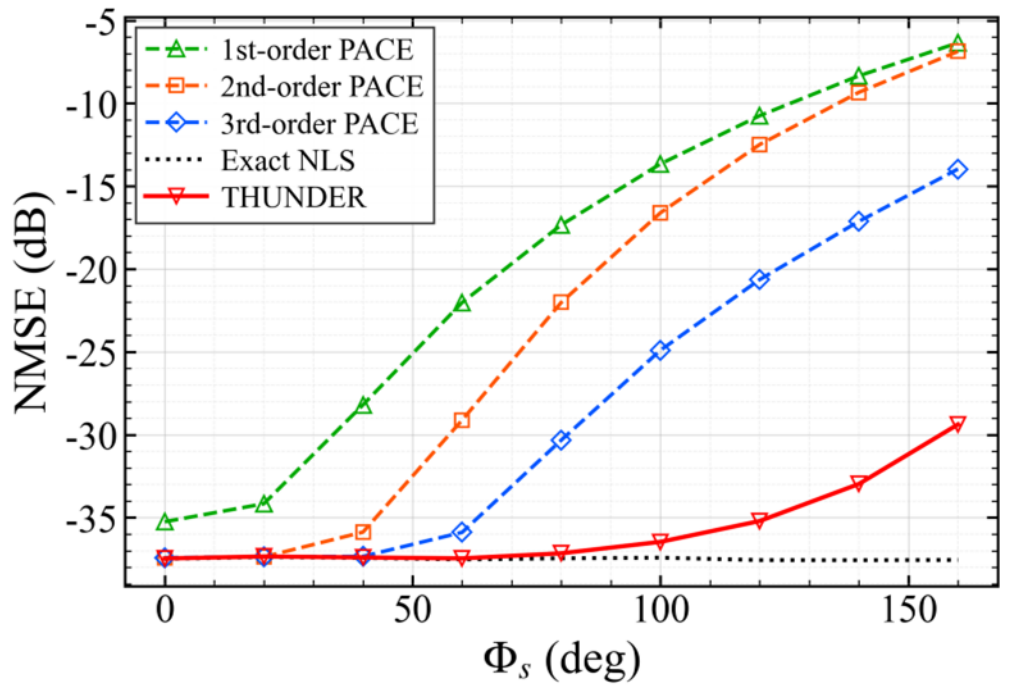

Fig. 5. Approximation-order comparison at SNR = 30 dB: NMSE versus pilot phase span $\Phi_s$ for finite-order lifted PACE ($1^{st} \sim 3^{rd}$ order), exact-model NLS, and THUNDER.

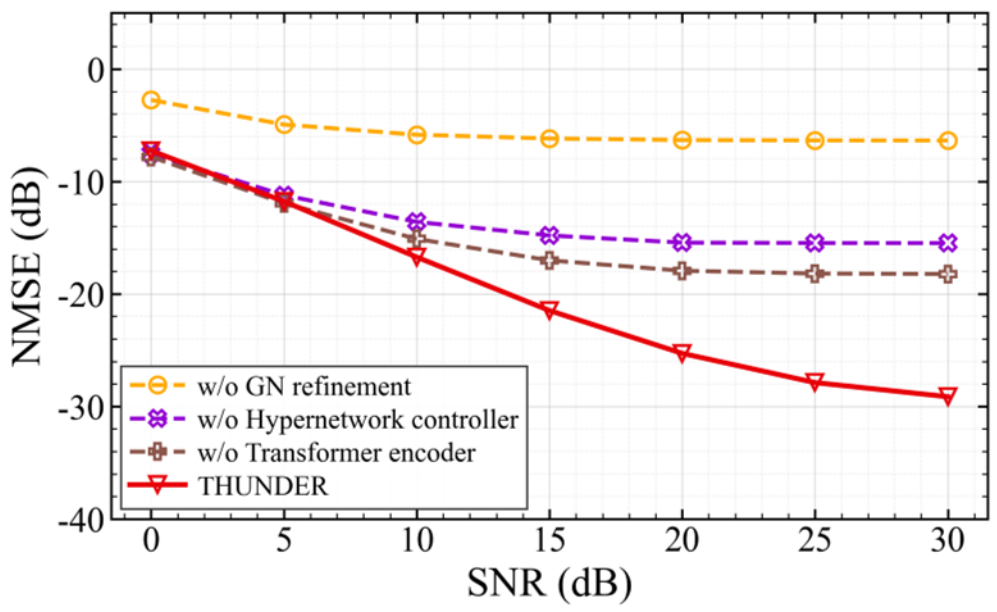

Fig. 6. Component ablation of THUNDER.

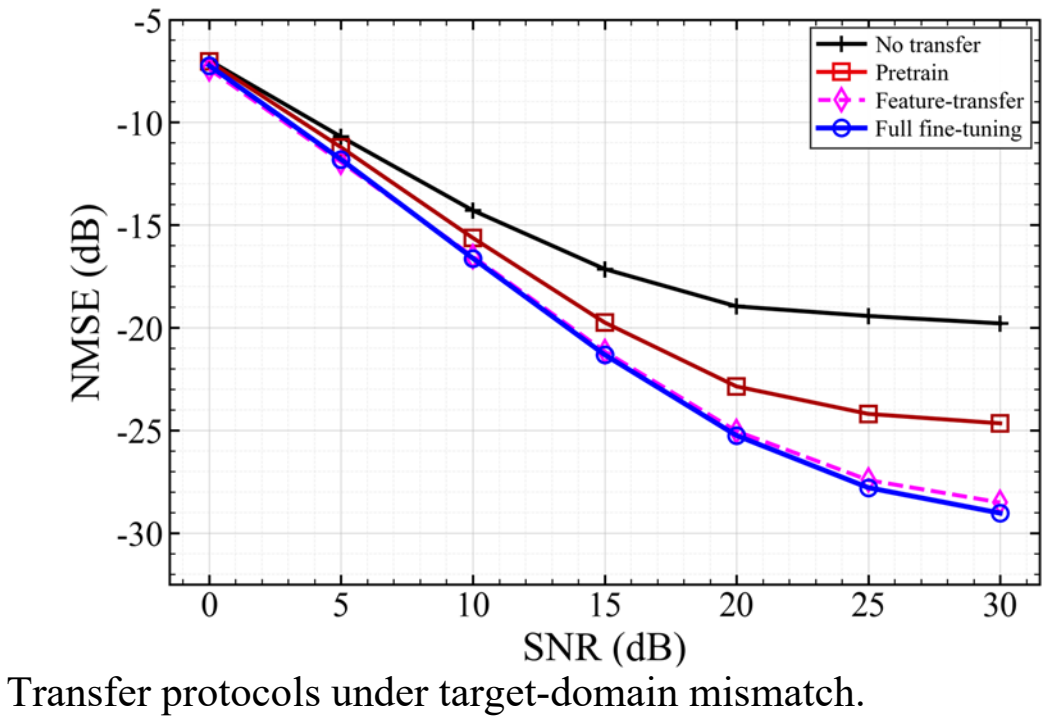

Fig. 7. Transfer protocols under target-domain mismatch.

The component ablation in Fig. 6 identifies the estimator structure responsible for the gain. Disabling the exact-model GN refinement reduces the estimator to the PACE initializer and restores the high-SNR floor, increasing the NMSE by about 22 dB and identifying exact-model refinement as the dominant mechanism. Disabling either the transformer encoder or the hypernetwork controller causes a smaller but consistent loss of 11–13 dB at high SNR, showing that exact-model refinement benefits from both pilot-wide context and frame-dependent control. These results indicate that the gain is not obtained by GN refinement alone, but by conditioning the exact-model recursion on the received pilot block.

The transfer comparison in Fig. 7 evaluates adaptation under source-to-target domain mismatch. The target-only model remains about 9 dB worse than full fine-tuning at high SNR, showing that the gain is not due to the target-data budget alone. Source pretraining improves over target-only training but still underperforms the transfer-adapted models, indicating that target-domain adjustment is needed under the noise-correlation mismatch. Feature-extraction transfer comes within about 0.5 dB of full fine-tuning over the high-SNR region. This small gap indicates that the transformer encoder provides a reusable pilot-context representation across the tested domains, whereas most target-domain adjustment is handled by the hypernetwork and reliability head. The transfer result therefore shows that control-module adaptation recovers most of the transfer gain, while the dominant accuracy improvement remains attributable to exact-model refinement.

## V. Conclusion

This paper proposed THUNDER, which combines closed-form PACE initialization with transformer-hypernetwork-controlled deep-unfolded GN refinement on the exact phase-exponential pilot model. By retaining the model-based channel-and-phase state while adapting the GN controls and pilot weights from pilot context, THUNDER suppresses the deterministic first-order error floor caused by residual phase drift. With a fixed five-step refinement, it achieves an 8.9 dB NMSE gain over the strongest learned baseline in the high-SNR and large-phase-span regime. The transfer result further shows that most target-domain adaptation can be handled by the hypernetwork and reliability head without retraining the transformer encoder.